# Neighbor Cell List Optimization for Femtocell-to-Femtocell Handover in Dense Femtocellular Networks

Mostafa Zaman Chowdhury, Bui Minh Trung, and Yeong Min Jang
Department of Electronics Engineering, Kookmin University, Seoul 136-702, Korea
E-mail: {mzceee, yjang}@kookmin.ac.kr

***Abstract*— Dense femtocells are the ultimate goal of the femtocellular network deployment. Among three types of handovers: femtocell-to-macrocell, macrocell-to-femtocell, and femtocell-to-femtocell, the latter two are the main concern for the dense femtocellular network deployment. For these handover cases, minimum as well appropriate neighbor cell list is the key element for the successful handover. In this paper, we propose an algorithm to make minimum but appropriate number of neighbor femtocell list for the femtocell-to-femtocell handover. Our algorithm considers received signal level from femto APs (FAPs); open and close access cases; and detected frequency from the neighbor femtocells. The simulation results show that the proposed scheme is able to attain minimum but optimal number of neighbor femtocell list for the possible femtocell-to-femtocell handover.**

## I. Introduction

The femtocellular networks [1]-[5], one of the most promising technologies to meet the demand of the tremendous increasing wireless capacity by various wireless applications for the future wireless communications. Among many advantages of femtocellular networks, the most important advantages are the offloading huge traffic from the expensive cellular networks to femtocellular networks; very small deployment cost; and use of the same frequency like cellular networks. Thus, the deployment of femtocells in the large scale [4], [5] is the ultimate goal for this technology.

The large and dense scale deployment of femtocells suffers from several challenges [3], [4], [6]. The handover is one challenge among several challenges. Three types of handovers may occur in dense femtocells environment; macrocell-to-femtocell handover, femtocell-to-macrocell handover, and femtocell-to-femtocell handover. Femtocell-to-macrocell handover does not suffer from additional challenges. However, the macrocell-to-femtocell and femtocell-to-macrocell handovers face some difficulties including the selection of appropriate femtocell for handover and the optimal neighbor femtocell list for the handover. In this paper we address the neighbor femtocell list problem for the femtocell-to-femtocell handover case.

In a dense femtocellular network deployment, a lot of femtocells are deployed within small coverage area. As a result, there may present huge interference effects. Whenever a mobile station (MS) realizes that the receive signal from the serving femto AP (FAP) is going down, the MS receive many signals from the several neighbor FAPs for the handover. Figure 1 shows that a MS in dense femtocellular deployment case may receive signals from many neighbor FAPs. Thus, the neighbor femtocell list based on the received signal will contain a large number of femtocells. Also there may have some hidden FAPs problem. The hidden FAP problem is the case when a neighbor FAP is very near to the MS but the MS cannot receive the signal due to some barrier (e.g., wall) between the MS and that FAP. Thus, the hidden FAPs will be out of neighbor femtocell list if the neighbor femtocell list is designed based on the receive signals only.

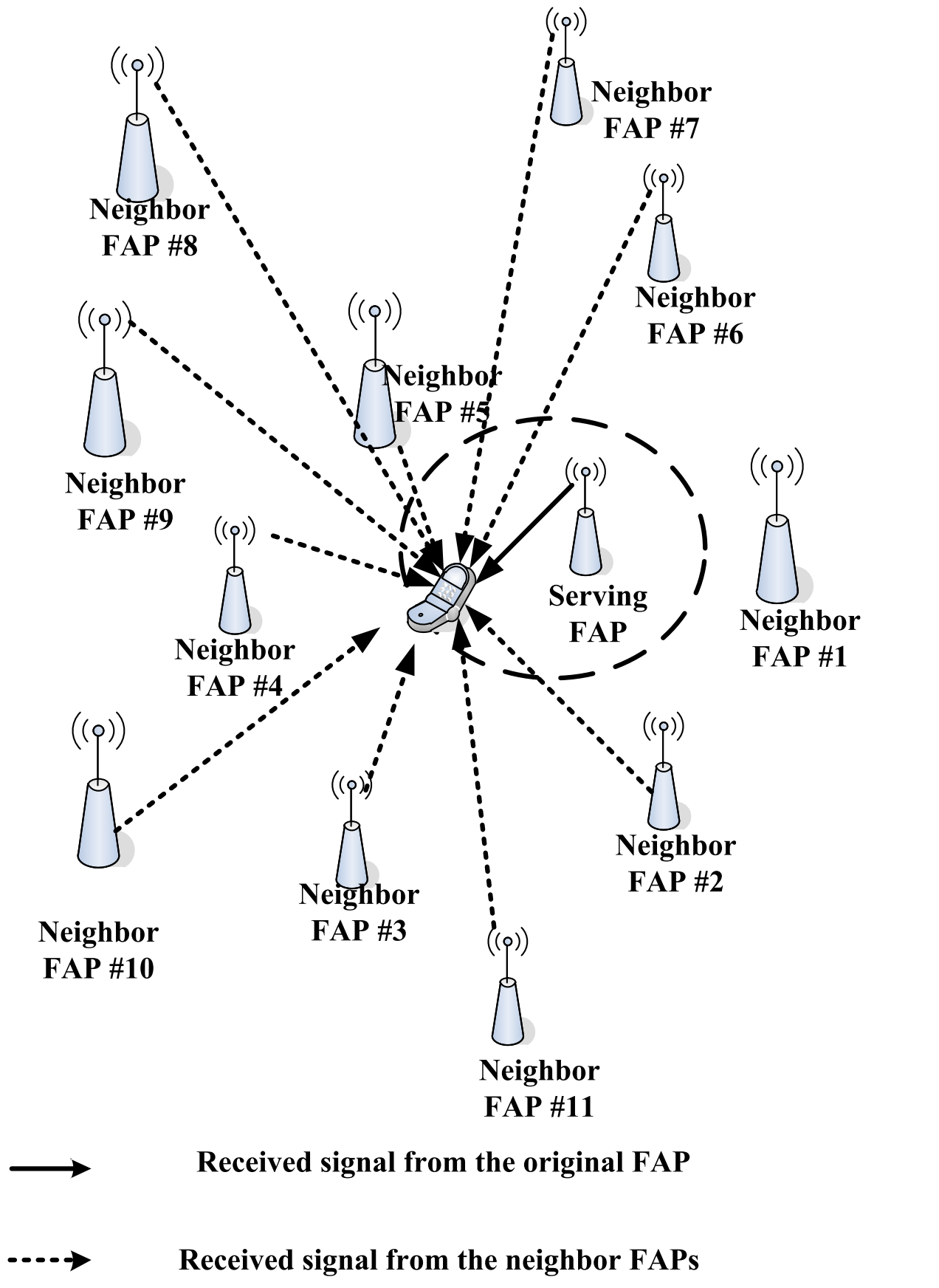

**Figure 1.** Femtocells scenario in dense femtocellular network deployment

Hence, two major challenges arise in the making of neighbor femtocell list based on the received signal level only; inclusion of some unnecessary femtocells in the neighbor femtocell list and exclusion of some important hidden FAPs from the neighbor femtocell list. In this paper we consider received signal level RSSI, frequency used by the serving FAP and the neighbor FAPs, and the location information for the optimal neighbor femtocell list in femtocell-to-femtocell handover.

The rest of this paper is organized as follows. Section II introduces the system model to support the dense femtocells. The neighbor cell list optimization techniques are described in Section III. In Section IV, we presented and compared the simulation results. Finally, we concluded our work in Section V.

## II. System Model to Support Dense Femtocells

Usual macrocellular networks utilize centralized RNC to control their associated macro base stations (BSs). One RNC is in charge of radio resource management of approximately 100 macro BSs. Femtocells are deployed within a macrocell coverage area or within a separate zone [5], [7]. Within one macrocell coverage area, thousands or tens of thousands of femtocells may exist. Thus, a single RNC needs the ability to control such large number of femtocells. It's not possible to handle so many FAPs using the existing network control entities. Therefore, for an efficient dense femtocells deployment, some addition features like self-organizing network (SON) capability and FAP management system should be added to the femtocellular networks compared to the traditional cellular BS. Figure 2 shows the femtocell system functional architecture for the concentrator-based femtocell/macrocell integrated network to support dense femtocellular networks. The FAP Management System (FMS) functionalities include configuration, fault detection, fault management, monitoring, and software upgrades. The registration part is used for the purpose of user equipment (UE) and FAPs registration. It maintains an authorized user list and permits only authorized user's access to a specific FAP. There are two kinds of registrations: FAP registration and UE registration. The IP network controller (INC) interfaces with the AAA proxy/server for provisioning of the FAP related information and service access control. The security gateway (SeGW) provides the mutual authentication, encryption, and data integrity for signalling, voice, and data traffics. A centralised reference clock with the FGW is used for synchronization purposes. The FAPs are controlled by a FAP controller (FAP-C). The database (DB) server in the macrocellular BS stores the information about the FAP's location and authorized cell list located in macrocell coverage area. Whenever a FAP is installed, the respective femto gateway (FGW) [4] provides FAP's position and its authorized user list to the macro Base Station DB server through CN. The Radio Network Subsystem (RNS) of FAP or macro BS controls the radio resources.

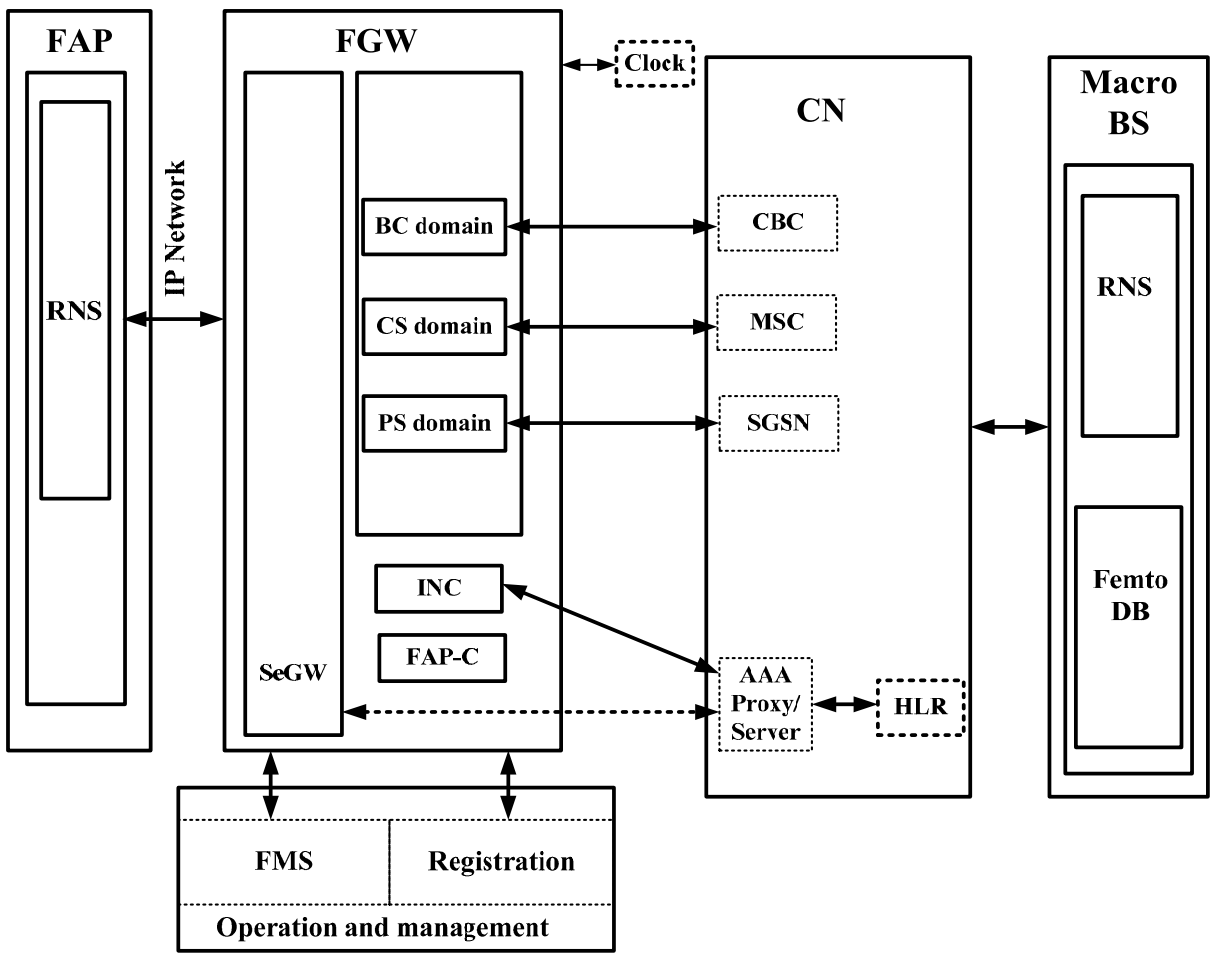

**Figure 2.** Femtocell system functional architecture for the concentrator-based femtocell/macrocell integrated network to support dense femtocellular networks

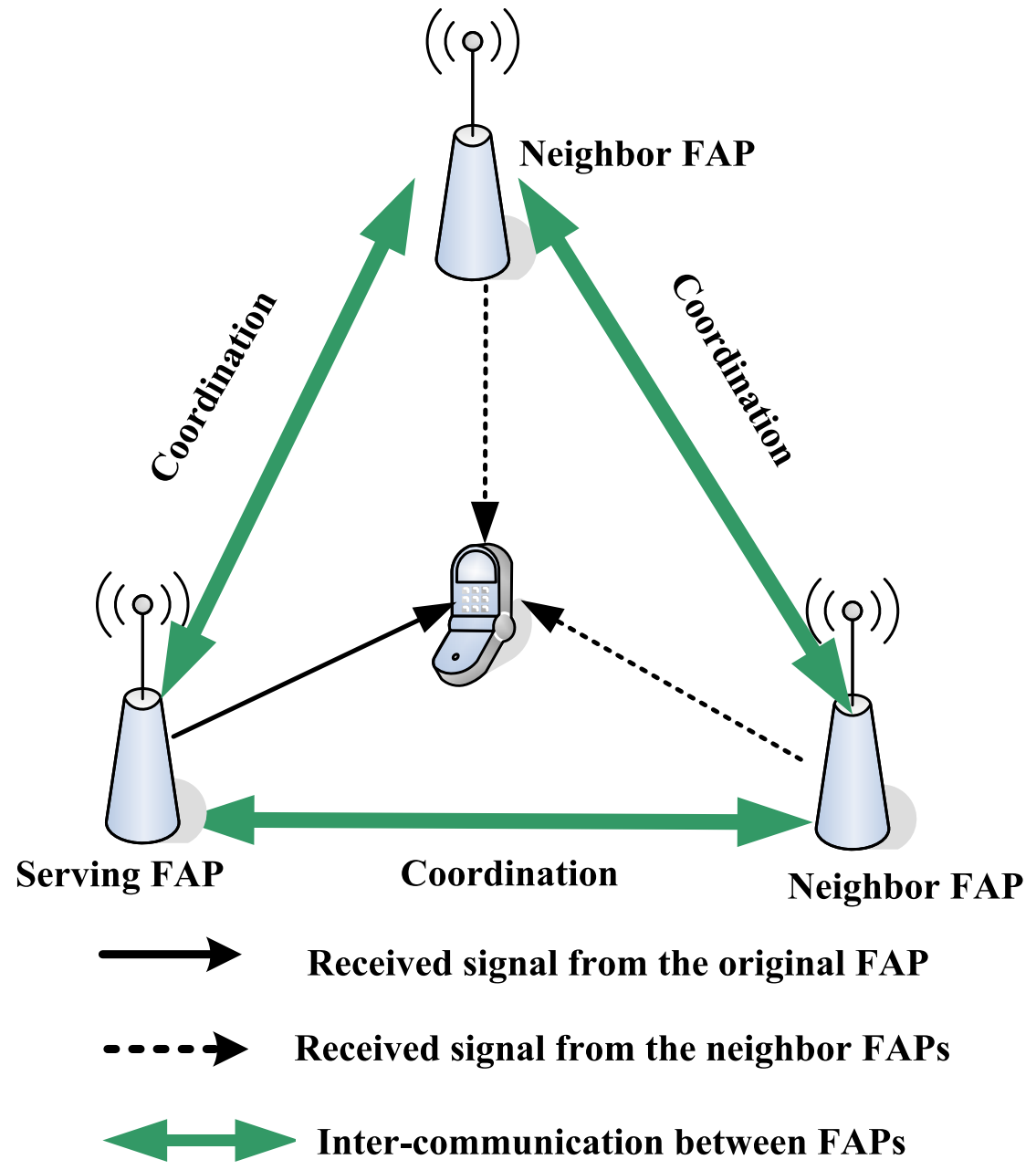

**Figure 3.** Coordination among the neighbor FAPs for the location update purposes.

Figure 3 shows the coordination among various FAPs to implement SON features (self-configuration, self-optimization, and self-healing) [4]. The location information can be exchanged among the neighbour FAPs for the building of an optimized neighbour femtocell list.

## III. Optimized Neighbor Femtocell List

Finding the neighboring FAPs and determining the appropriate FAP for handover are challenging for optimum handover decision [4]. Handover from the femtocell-to-femtocell in dense femtocellular network environment suffers some additional challenges because of dense neighbor femtocells. In this handover, MS needs to select the appropriate target FAP among many FAPs. The femtocell-to-femtocell handover scheme creates a problem if there is no minimum number of femtocells in the neighbor femtocell list. The MSs use much more power consumption for scanning many FAPs, and the MAC overhead becomes significant. This increased size of neighbor FAP list message and broadcasting of large information occurs too much overhead. So, an appropriate and optimal neighbor FAP list is essential.

Whenever an MS moves away from its serving FAP, the MS detects many neighbor FAPs due to dense deployment of femtocells as well as the MS detects the presence of macrocell coverage. The FAPs coordinate with each other to facilitate a smooth handover. If a large number of FAPs are deployed in an indoor building or femto zone area, signals from different FAPs will interfere with each other. Thus during the handover phase it is quite difficult to sense the actual FAP for which the user is going to handover. The need of minimum neighbor femtocell list is essential to make minimum number of scanning and signal flowing during the handover. Large neighbor femtocell list causes many unnecessary scanning for the handover. Also missing of some hidden femtocells in the neighbor femtocell list causes the failure of handover. Our main goal is to build such a neighbor femtocell list for the femtocell-to-femtocell handover so that the list contains minimum number of femtocells. However, the list includes some hidden femtocells. Figure 4 shows a scenario of dense femtocellular network deployment where several FAPs are situated as a neighbor femtocell.

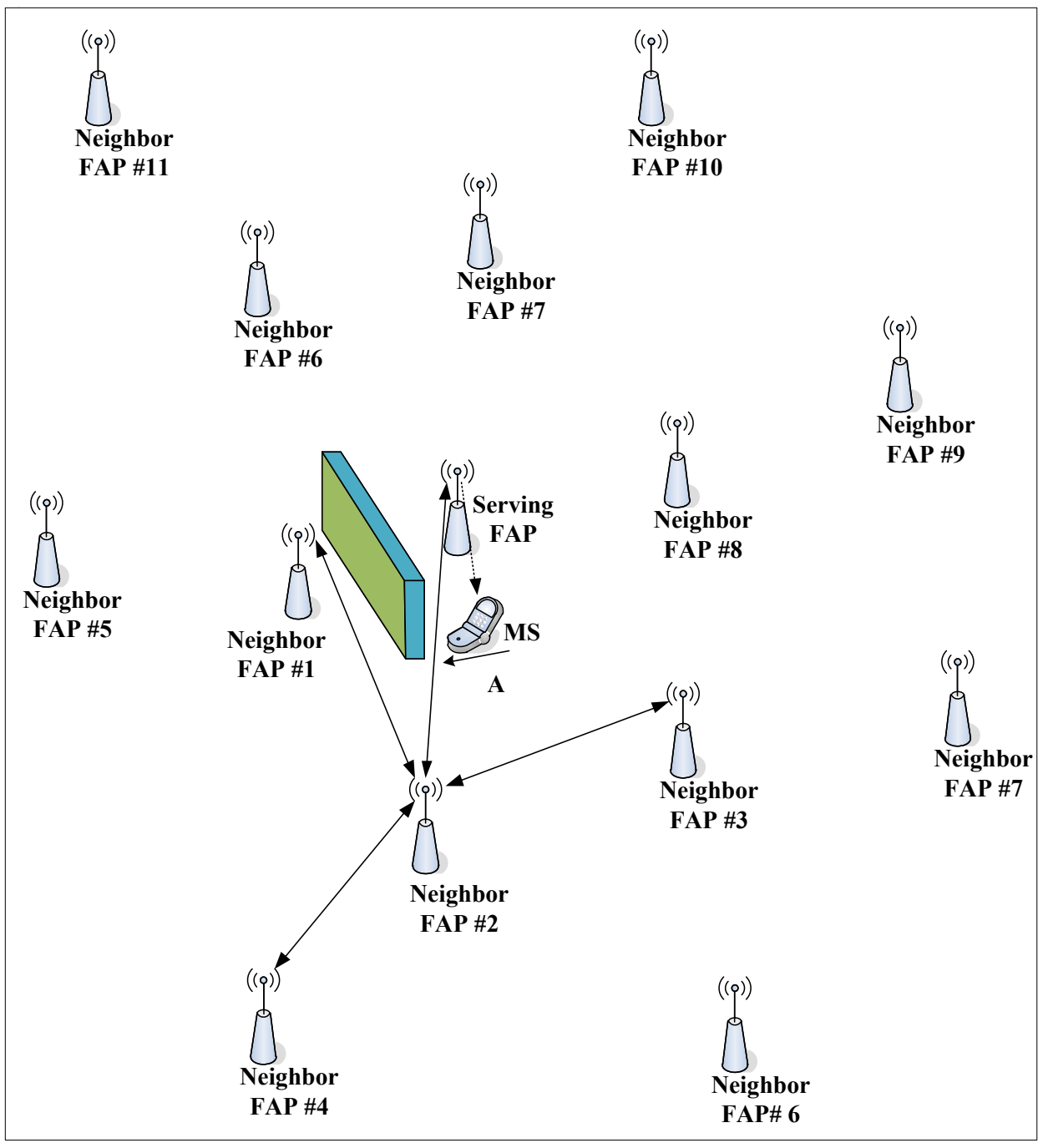

**Figure 4.** A scenario of dense femtocellular network deployment where several hidden FAPs and other FAPs are situated as a neighbor femtocell.

In Figure 4, for the MS at position "A" shown here, the optimized neighbor femtocell list must contain FAP #1, FAP #2, FAP #3, and FAP #8. However, due a wall and other obstacle between the MS and the FAP #1, the MS cannot receive sufficient signal from this FAP. Serving FAP and FAP #1 also cannot coordinate each other. Thus a neighbor femtocell list only based on the RSSI (received signal strength indicator) measurement cannot include FAP #1 in the neighbor femtocell list. In this situation, FAP #2 can give the location information of FAP #1. Thus, getting this location information, the neighbor femtocell list includes FAP #1.

Figure 5 shows the flow mechanism for the design of the optimal neighbor femtocells list. Our proposed scheme considers initially the received RSSI level. For the dense femtocellular network deployment, the frequency for each FAP is allocated based on the neighbor overlapping femtocells. Thus, the overlapping two femtocells do not use same frequency [5]. Only same frequency is used by apart femtocells. Thus, initially we deduct those FAPs from the neighbor femtocell list that uses same frequency as the serving FAP. Finally we added the hidden femtocells in the neighbor femtocell list using the location information coordinated by neighbor FAPs.

The FAP that are initially listed as the neighbor femtocell list based on the received RSSI level can be expressed as a set $A$:

$$A = \{ \cdots FAP \ \#i(RSSI_i), \cdots : 1 \le i, \ RSSI_i \ge S_{T0} \} \quad (1)$$

where FAP #i($RSSI_i$) represents that *i-th* neighbor FAP from which the received RSSI level by the MS is greater than or equal to $S_{T0}$. $S_{T0}$ is the minimum level of received signal from a FAP that can be detected by a MS.

Instead of considering only the RSSI level, we consider RSSI level; frequency used by the serving FAP and *i-th* neighbor FAP; and the location information to construct an appropriate neighbor femtocell list.

We consider little higher RSSI level $S_{T1}$, compared to $S_{T0}$, to select better signal quality FAPs. Some hidden femtocells are picked for the neighbor femtocell list from where the received signal is less than $S_{T1}$. The FAPs listed for the neighbor femtocell list based on the threshold level $S_{T1}$ can be expressed as set B:

$$B = \{ \cdots FAP \ \#j(RSSI_j), \cdots : 1 \le j, \ RSSI_j \ge S_{T1} \} \quad (2)$$

In the dense femtocell deployment, same frequency is not used for the overlapped femtocells [4], [5]. Therefore same frequencies are used by two femtocells those are little far away. So from the neighbor femtocell list we deduct those femtocells which use same frequency as serving femtocells. The femtocells those can be categorized in this group can be expressed as set C:

$$C = \{ \cdots FAP \ \#k(f_k), \cdots : C \in B, f_s \cup f_i = f_s \} \quad (3)$$

where FAP #k($f_k$) represents that *k-th* neighbor femtocell that use frequency $f_k$. Whereas $f_s$ is the frequency used by the serving femtocell.

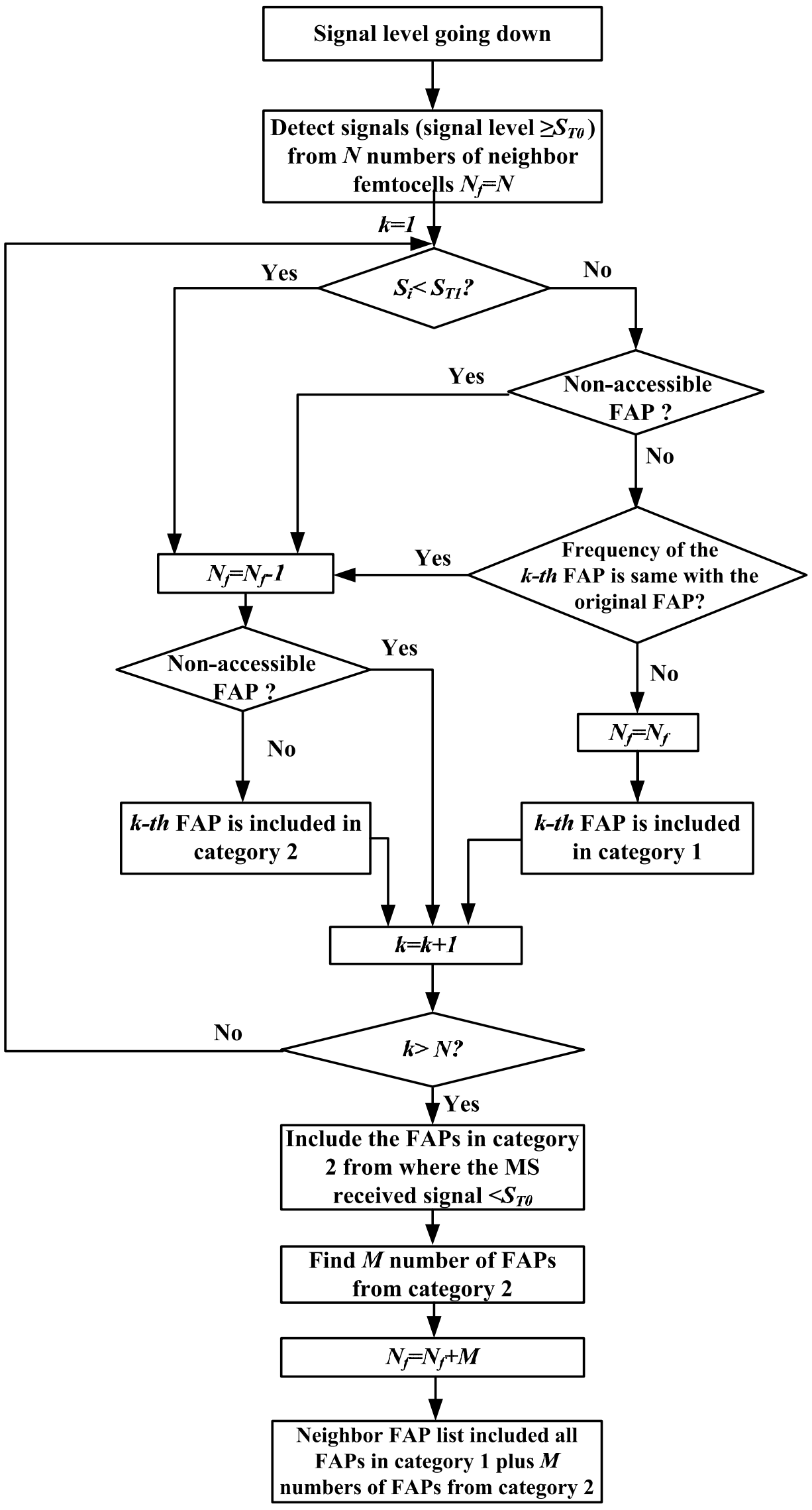

**Figure 5.** The flow mechanism for the design of the optimal neighbor femtocells list during femtocell-to-femtocell handover.

Now, we apply the location information for the neighbor femtocell list to include the hidden FAPs in the neighbor femtocell list. The hidden femtocells are chosen from the category 2 femtocells. The femtocells are in this category included (a) the femtocells from where the received RSSI level is less than $S_{T1}$, or (b) which femtocells use the same frequency as the serving femtocell. As the serving FAP can coordinate with some nearest FAPs, [4], [5] then nearest FAPs can inform the location of some hidden FAPs. Thus the hidden FAPs within a range of distance can be included in the neighbor femtocell list. The femtocells those are included in this group can be expressed as set D:

$$D = \{\cdots FAP\#m(RSSI_m, f_m, d), \cdots : (RSSI_m < S_{T1} \text{ or } f_s \cup f_m = f_s) \text{ and } d \leq \Gamma\} \quad (4)$$

where, $d$ is the distance between the MS and the $m$-$th$ neighbor femtocell that use frequency $f_m$. The $m$-$th$ femtocell will be included in this group only if the distance between the MS and the $m$-$th$ neighbor femtocell is less than or equal to a predefined threshold distance $\Gamma$.

Considering the above three facts (RSSI level, frequency, and location information) the femtocells included in the final neighbor femtocell list are:

$$E = (B / C) \cup D \quad (5)$$

## IV. Performance Analysis

We verified the performance of the proposed optimized neighbor femtocell list scheme using simulation result. Table 1 shows the basic simulation parameters. We randomly generate the location of femtocells with respect to a reference femtocell. We also assume random manner about the hidden femtocell. The reference MS is assumed at the edge of the reference femtocell. We consider both the open access and close access randomly in the simulation.

**Table 1.** Simulation assumptions

| | |
|---|---|
| Radius of femtocell coverage area | 10 m |
| Carrier frequency for femtocells | 1.8 GHz |
| Transmit signal power by macro BS | 1.5 W |
| Maximum transmit power by FAP | 10 mW |
| Propagation model for femtocell $(L_{femto})$ | $20\log_{10}f + N\log_{10}d + L_f(n)$ -28 dB |
| Height of FAP | 2 m |
| Detected value of received signal from original FAP $(S_{T0})$ | -90 dBm |
| Threshold value of received signal from a neighbor FAP $(S_{T1})$ | -75 dBm |

Figure 6 shows the probability comparison that the target femtocell is missing from the neighbor femtocell list. In a traditional neighbor femtocell list based on the received signal strength cannot include the hidden femtocells in the neighbor femtocell list. Thus, there is a possibility that the target femtocell is not included in the neighbor femtocell list. This causes a failure of the handover to the target femtocell. The increasing of number of deployed femtocells increases the possibility that a neighbor femtocell coordinates with the reference femtocell and keep informed about the location of the hidden neighbor femtocell. As a result, the increasing number of deployed femtocells causes the reduction of probability that the hidden femtocell is out of the neighbor femtocell list. Also missing the appropriate neighbor femtocell from the neighbor femtocell list also causes a handover failure. Thus the handover failure rate decreases for the proposed scheme with the increase of the number of deployed femtocells.

Figure 7 shows the comparison of the numbers of neighbor femtocells in the neighbor femtocell list for two schemes based on different parameters matrix. The result shows that the neighbor femtocell list contains very small number of femtocells during the femtocell-to-femtocell handover. Thus, the signal flow for the femtocell-to-femtocell handover process became very small.

The results in Figures 6 and 7 demonstrate that the neighbor femtocell list for the femtocell-to-femtocell handover contains optical number of femtocells. However, the reduced number of femtocells in the neighbor femtocell list does not increase the handover failure probability. Instead of increasing the handover failure probability due to hidden femtocell problem, our scheme reduces the handover failure probability.

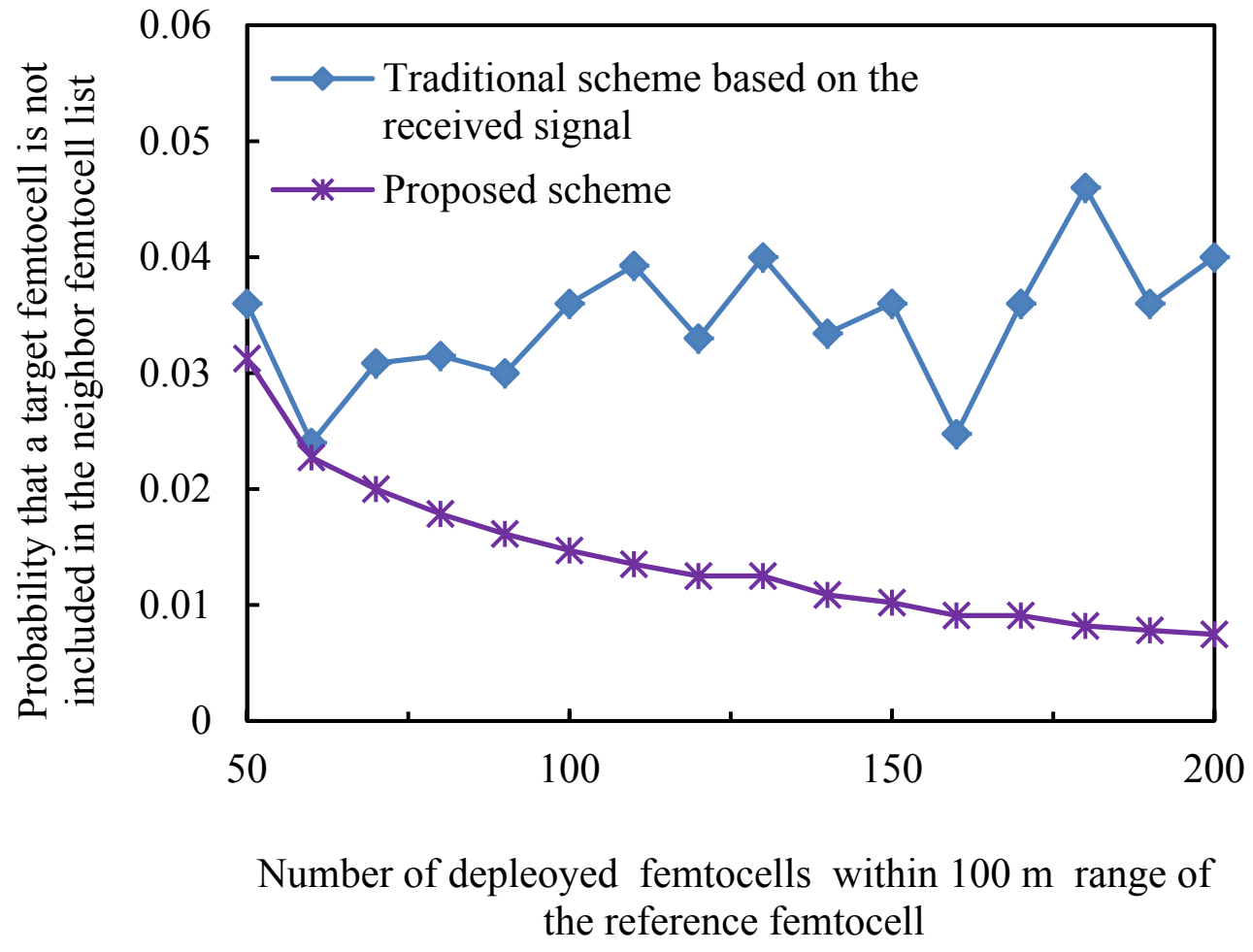

**Figure 6.** The probability comparison when the target femtocell is missing from the neighbor femtocell list.

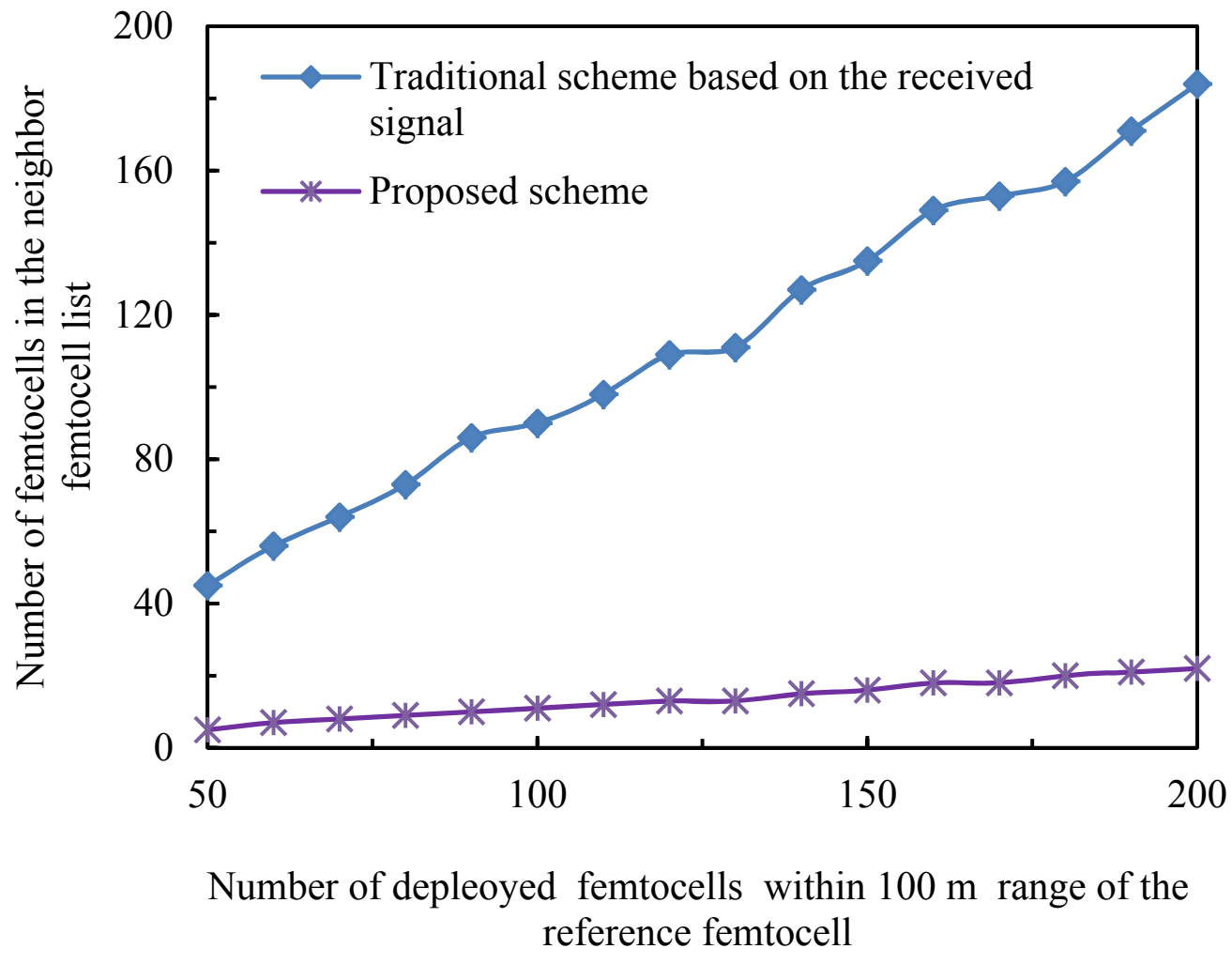

**Figure 7.** A comparison of the numbers of neighbor femtocells in the neighbor femtocell list for two schemes based on different parameters matrix.

## V. Conclusions

For the femtocell-to-femtocell handover in the dense femtocellular network deployment, the two most important considerations are the small number of neighbor femtocells in the neighbor femtocell list and including the hidden neighbor femtocells in the neighbor femtocell list. The small number of neighbor femtocells in the neighbor femtocells list reduces the power consumption for scanning many FAPs and also reduces the MAC overhead. The inclusion of the hidden neighbor femtocells reduces the femtocell-to-femtocell handover failure probability. We consider received signal level as well as the location information using SON capabilities of the FAPs for the neighbor femtocell list. We select femtocells from two categories. Some femtocells are categorized in first category from which the received signals are greater than or equal to a threshold level. The femtocells are grouped in second category from which the received signals are less than a threshold level or the serving FAP and the neighbor FAP use the same frequency. The hidden femtocells are listed from the second category.

The dense femtocell that is the ultimate goal of the femtocellular networks deployment and successful femtocell-to-femtocell handover is the key success parameter for the dense femtocellular network deployment. The results shown in this paper clearly indicate the advantages of our proposed scheme for the dense femtocellular network deployment.

## Acknowledgement

This work was supported by the IT R&D program of MKE/KEIT [10035362, Development of Home Network Technology based on LED-ID].